\documentclass[12pt]{iopart}
\usepackage{graphicx}
\begin{document}
\title[Feshbach resonances in ultracold atomic and molecular collisions]
{Feshbach resonances in ultracold atomic and molecular collisions:\\
Threshold behaviour and suppression of poles in scattering
lengths}
\author{Jeremy M. Hutson}
\address{Department of Chemistry, University of Durham, South Road,
Durham, DH1~3LE, England}

\date{\today}

\begin{abstract}
In the absence of inelastic scattering, Feshbach resonances
produce poles in scattering lengths and very large peaks in
elastic cross sections. However, inelastic scattering removes the
poles. Whenever the resonant state is coupled comparably to the
elastic and inelastic channels, the scattering length exhibits
only a small oscillation and peaks in cross sections are
significantly suppressed. A resonant scattering length is defined
to characterize the amplitude of the oscillation, and is shown to
be small for many collisions of ultracold molecules. The results
suggest that cross sections for some ultracold collision processes
will be much less sensitive to details of the potential than has
been expected.
\end{abstract}
\pacs{03.65.Nk,03.75.Nt,34.10.+x,34.50.-s,82.20.Xr}

\maketitle

\section{Introduction}

A Feshbach resonance \cite{Feshbach:1958} occurs when a bound
state of a 2-particle system lies above a dissociation threshold
and is coupled to the continuum. Collision properties show sharp
features (peaks and troughs) near the energy of the resonance. In
recent years, Feshbach resonances have come into prominence in the
study of ultracold atomic gases. In these systems the positions of
resonances can often be adjusted using applied magnetic fields,
and it is possible to {\it control} the interactions between atoms
by tuning resonances to near-zero collision energy
\cite{Timmermans:1999, Hutson:IRPC:2006, Koehler:RMP:2006}.
Magnetic tuning through Feshbach resonances has been used to
produce molecules in both bosonic and fermionic quantum gases.
Long-lived molecular Bose-Einstein condensates of fermion dimers
have been produced, and the first signatures of ultracold
triatomic and tetraatomic molecules have been observed. The new
capabilities in atomic physics have had important applications in
other areas: for example, the tunability of atomic interactions
has been used to explore the crossover between Bose-Einstein
condensation (BEC) and Bardeen-Cooper-Schrieffer (BCS) behaviour
in dilute gases. There is now great interest in extending the
capabilities from ultracold atomic to molecular systems, to
explore the properties of dipolar quantum gases and develop new
forms of quantum control.

Most interpretations of Feshbach resonances have used concepts
from the {\it two-channel model} \cite{Koehler:RMP:2006}, in which
the bound state and the continuum are each represented by one
scattering channel. This captures much of the crucial resonant
behaviour observed in ultracold atom-atom scattering. In
particular, it predicts that the scattering length passes through
a pole and the elastic scattering cross section exhibits a very
large peak at a zero-energy resonance. However, it is known from
early work on nuclear reactions \cite{Bethe:1937} that inelastic
processes suppress resonant peaks in cross sections. The purpose
of this paper is to explore the consequences of such effects for
ultracold atomic and molecular collisions. Whenever the resonant
state is coupled comparably to the incoming and inelastic
channels, the scattering length exhibits only a small oscillation
and the peaks in cross sections are dramatically suppressed. This
is particularly important for the prospect of controlling
molecular collisions.

This paper will first summarize the results of 2-channel resonance
theory, to define notation and establish a basis for comparison.
The major differences introduced by inelastic scattering will then
be considered. The results are general, but to assist
visualisation the equations will be illustrated with examples
taken from the elastic and inelastic scattering of NH molecules
with He \cite{Gonzalez-Martinez:2007}.

\section{Resonances in the absence of inelastic scattering}

When there is only a single open channel with orbital angular
momentum $l$, the long-range wavefunction may be written
\begin{equation}
\psi^{\rm open}(r) = N k^{-1/2} r^{-1} \sin [kr - l\pi/2 +
\delta(k)]
\end{equation}
where $\delta(k)$ is the phase shift and the wave vector $k$ is
defined in terms of the kinetic energy $E_{\rm kin}$ and reduced
mass $\mu$ by $E_{\rm kin}=\hbar^2 k^2/2\mu$.
In the ultracold regime, cross sections are dominated by s-wave
scattering, with $l=0$. The most important parameter is the
energy-dependent s-wave scattering length $a(k)$, defined by
\begin{equation}
a(k)= \frac{-\tan\delta(k)}{k}. \label{eqad}
\end{equation}
This becomes constant at limitingly low energy, with corrections
given by effective range theory \cite{Hickelmann:1971},
\begin{equation}
a(k) = a(0) + \frac{1}{2}k^2 r_0 a(0)^2 + {\cal O}(k^4),
\label{eqeffr}
\end{equation}
where $r_0$ is the effective range. The elastic cross section is
given exactly in terms of $a(k)$ by
\begin{equation}
\sigma_{\rm el}(k) = \frac{4\pi a^2}{1+k^2 a^2}. \label{eqsiga}
\end{equation}
For collisions of identical bosons, the factor of 4 is replaced by
8. However, the present work will omit such extra factors of 2.

If there is only one open channel, the behaviour of the phase
shift $\delta$ is sufficient to characterize a resonance. It
follows a Breit-Wigner form as a function of energy,
\begin{equation}
\delta(E) = \delta_{\rm bg} + \tan^{-1}
\left[\frac{\Gamma_E}{2(E_{\rm res}-E)}\right], \label{eqbw}
\end{equation}
where $\delta_{\rm bg}$ is a slowly varying background term,
$E_{\rm res}$ is the resonance position and $\Gamma_E$ is its
width (in energy space). The phase shift thus increases sharply by
$\pi$ across the width of the resonance. In general the parameters
$\delta_{\rm bg}$, $E_{\rm res}$ and $\Gamma_E$ are weak functions
of energy, but this is neglected in the present work apart from
threshold behaviour.

As a function of magnetic field at constant $E_{\rm kin}$, the
phase shift follows a form similar to Eq.\ \ref{eqbw},
\begin{equation}
\delta(B) = \delta_{\rm bg} + \tan^{-1}
\left[\frac{\Gamma_B}{2(B_{\rm res}-B)}\right], \label{eqbwb}
\end{equation}
where $B_{\rm res}$ is the field at which $E_{\rm res}=E=E_{\rm
thresh}+E_{\rm kin}$. The width $\Gamma_B$ is a signed quantity
given by $\Gamma_B=\Gamma_E/\Delta\mu$, where the magnetic moment
difference $\Delta\mu$ is the rate at which the energy $E_{\rm
thresh}$ of the open-channel threshold tunes with respect to the
resonance energy,
\begin{equation}
\Delta\mu = \frac{dE_{\rm thresh}}{dB}-\frac{dE_{\rm res}}{dB}.
\end{equation}
$\Gamma_B$ is thus negative if the bound state tunes upwards
through the energy of interest.

\begin{figure}[tb]
\includegraphics[width=63mm]{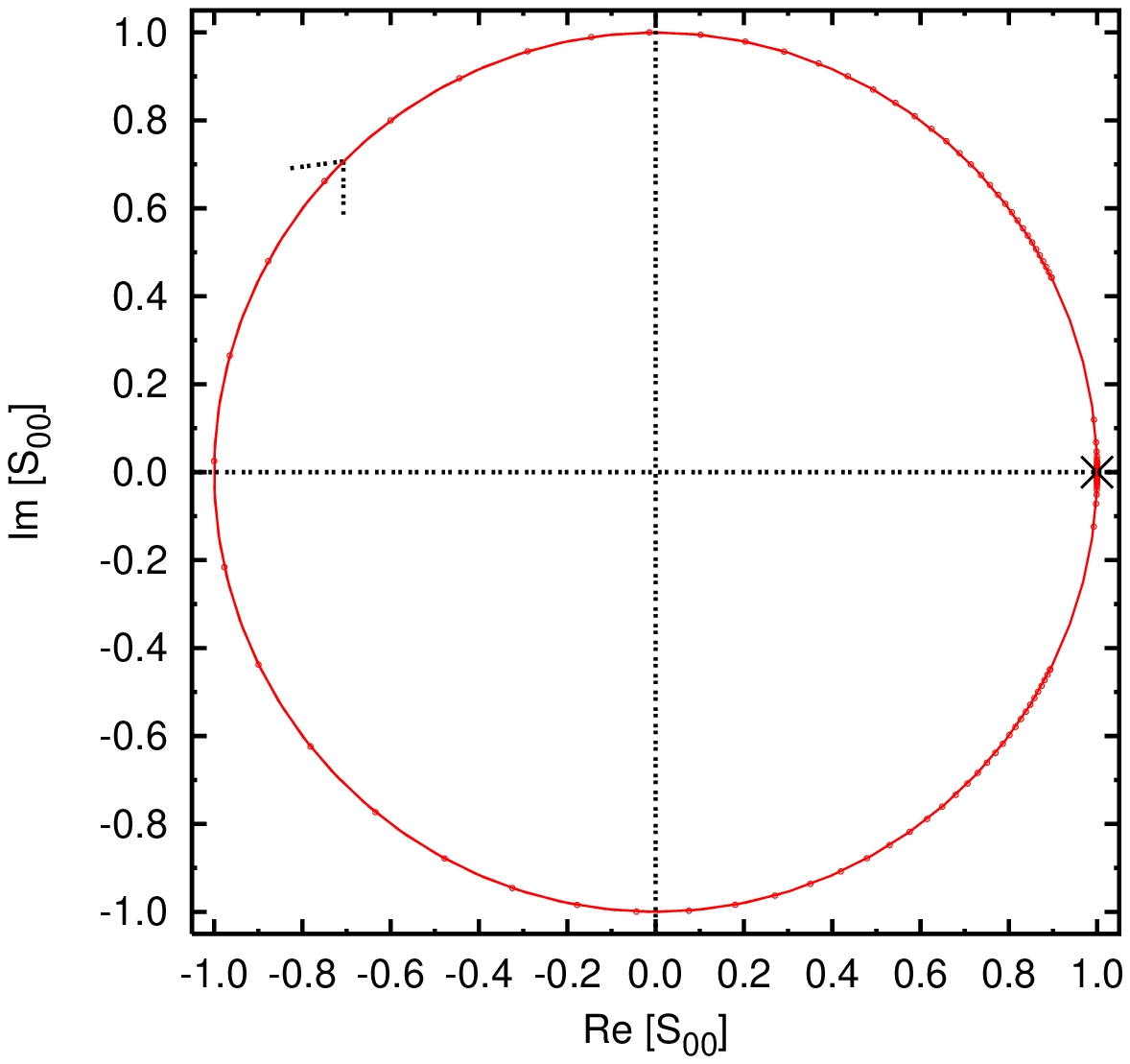}
\includegraphics[width=69mm]{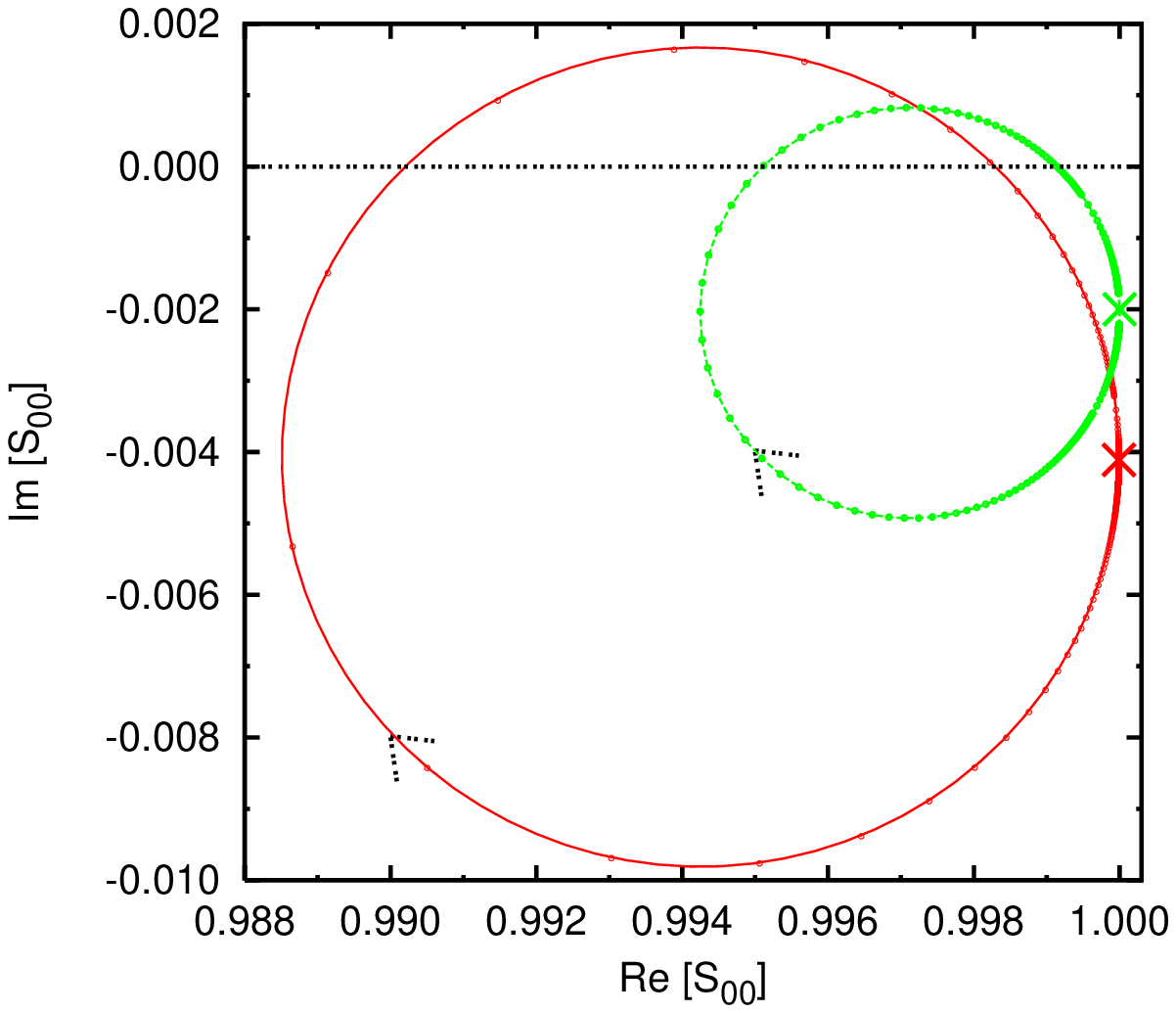}
\caption{The resonant circles described by $S$ matrix elements for
low-energy elastic scattering for two different resonances in He +
NH ($^3\Sigma^-$). Left panel: the circle of radius 1 when only
elastic scattering is allowed: incoming channel $n=0$, $m_s=-1$ at
$E_{\rm kin}=10^{-6}$ K. Right panel: the much smaller circles
(note the different scale) when both elastic and inelastic
scattering are allowed: incoming channel $n=0$, $m_s=0$ at $E_{\rm
kin}=10^{-6}$~K (green, smaller circle) and $4 \times 10^{-6}$~K
(red, larger circle). The crosses show values far from resonance.
In both cases the resonant state has $n=0$, $m_s=+1$.}
\label{figsm}
\end{figure}

Across an elastic scattering resonance, the $S$ matrix element
$S=e^{2{\rm i}\delta}$ describes a circle of radius 1 in the
complex plane as a function of either energy or magnetic field, as
shown in the left panel of Figure \ref{figsm}. In the ultracold
regime, the background phase shift $\delta_{\rm bg}$ goes to zero
as $k\rightarrow0$ according to Eq.\ \ref{eqad} (with $a_{\rm bg}$
constant and finite), but the resonant term still exists. The
scattering length passes through a pole when
$\delta=\left(n+\frac{1}{2}\right)\pi$, corresponding to $S=-1$.
The scattering length follows the formula \cite{Moerdijk:1995},
\begin{equation}
a(B) = a_{\rm bg} \left[ 1 - \frac{\Delta_B}{B - B_{\rm res}}
\right]. \label{eqares}
\end{equation}
The elastic cross section given by Eq.\ \ref{eqsiga} thus shows a
sharp peak of height $4\pi/k^2$ at resonance. The two widths
$\Gamma_B$ and $\Delta_B$ are related by
\begin{equation}
\Gamma_B=-2a_{\rm bg} k \Delta_B. \label{eqgamdel}
\end{equation}
At limitingly low energy, $\Gamma_B$ is proportional to $k$
\cite{Timmermans:1999} while $\Delta_B$ is constant.

\section{Resonances in the presence of inelastic scattering}

In the presence of inelastic collisions, the scattering matrix has
elements $S_{ii'}$. The diagonal S-matrix element in the incoming
channel $0$ has magnitude $S_{00}\le 1$ and may be written in
terms of a complex phase shift $\delta_0$ with a positive
imaginary part \cite{Mott:p380:1965},
\begin{equation}
S_{00}(k_0)=e^{2{\rm i}\delta_0(k_0)}, \label{eqsd}
\end{equation}
where $k_0$ is the wave vector in the incoming channel. This can
be expressed in terms of a complex energy-dependent scattering
length, $a(k_0)=\alpha(k_0)-{\rm i}\beta(k_0)$ \cite{Bohn:1997,
Balakrishnan:scat-len:1997}, defined by analogy with Eq.\
\ref{eqad} as
\begin{equation}
a(k_0)= \frac{-\tan\delta_0(k_0)}{k_0} = \frac{1}{{\rm i}k_0}
\left(\frac{1-S_{00}(k_0)}{1+S_{00}(k_0)}\right). \label{eqacomp}
\end{equation}
$a(k_0)$ again becomes constant at limitingly low energy. The
elastic and total inelastic cross sections are exactly
\cite{Cvitas:li3:2007}
\begin{equation}
\sigma_{\rm el}(k_0) = \frac{4\pi|a|^2}{1+k_0^2|a|^2+2k_0\beta}
\label{eqsigela}
\end{equation}
and
\begin{equation}
\sigma_{\rm inel}^{\rm tot}(k_0) =
\frac{4\pi\beta}{k_0(1+k_0^2|a|^2+2k_0\beta)}. \label{eqsiginela}
\end{equation}

When there are several open channels, the quantity that follows
the Breit-Wigner form (\ref{eqbw}) or (\ref{eqbwb}) is the
S-matrix eigenphase sum \cite{Hazi:1979, Ashton:1983}, which is
the sum of phases of the {\it eigenvalues} of the $S$ matrix. The
eigenphases and the eigenphase sum are real, unlike the phases
$\delta_i$ obtained from individual diagonal elements, because the
$S$ matrix is unitary, so that all its eigenvalues have modulus 1.

Across a resonance, the individual $S$ matrix elements describe
circles in the complex plane \cite{Brenig:1959, Taylor:p411:1972},
\begin{equation}
S_{ii'}(E) = S_{{\rm bg,}ii'} - \frac{{\rm i} g_{Ei} g_{Ei'}}{E -
E_{\rm res} + {\rm i}\Gamma_E/2}, \label{eqsiie}
\end{equation}
where $g_{Ei}$ is complex. The radius of the circle in $S_{ii'}$
is $|g_{Ei} g_{Ei'}|/\Gamma_E$. The {\it partial width} for
channel $i$ is usually defined as a real quantity,
$\Gamma_{Ei}=|g_{Ei}|^2$, but here we also need a corresponding
phase $\phi_i$ to describe the {\it direction} of the circle in
the complex plane, $g_{Ei}^2=\Gamma_{Ei} e^{2{\rm i}{\phi_i}}$.
For a narrow resonance, the total width is just the sum of the
partial widths,
\begin{eqnarray} \Gamma_E = \sum_i \Gamma_{Ei}.
\end{eqnarray}

As a function of magnetic field at constant $E_{\rm kin}$,
\begin{equation}
S_{ii'}(B) = S_{{\rm bg,}ii'} - \frac{{\rm i} g_{Bi} g_{Bi'}}{B -
B_{\rm res} + {\rm i}\Gamma_B/2}, \label{eqsii}
\end{equation}
where $g_{Bi} = g_{Ei}/\Delta\mu^{1/2}$ and the width $\Gamma_B$
and partial widths $\Gamma_{Bi}$ are signed quantities,
$\Gamma_B=\Gamma_E/\Delta\mu$ and
$\Gamma_{Bi}=\Gamma_{Ei}/\Delta\mu$.


The partial widths for {\it elastic} channels (degenerate with the
incoming channel) are proportional to $k_0$ at low energy. We may
define a reduced partial width $\gamma_{E0}$ or $\gamma_{B0}$ for
the incoming channel by
\begin{eqnarray}
\Gamma_{E0}(k_0) = 2k_0\gamma_{E0} \quad\hbox{or}\quad
\Gamma_{B0}(k_0) &=& 2k_0\gamma_{B0}, \label{eqgamlin}
\end{eqnarray}
and the reduced widths are independent of $k_0$ at low energy. By
contrast, the partial widths for {\it inelastic} channels depend
on open-channel wavefunctions with large wave vectors $k_i$ and
are effectively independent of $k_0$ in the ultracold regime. If
the inelastic partial widths $\Gamma_{Ei}$ (or $\Gamma_{Bi}$) are
non-zero, they eventually dominate $\Gamma_{E0}$ (or
$\Gamma_{B0}$) as $k_0$ decreases. The radius of the circle
(\ref{eqsii}) described by $S_{00}$ thus drops linearly to zero as
$k_0$ decreases, as shown in the right panel of Figure
\ref{figsm}. This is {\it qualitatively different} from the
behaviour in the absence of inelastic channels.

As a function of magnetic field, the scattering length passes
through a pole only if $\delta_0$ passes through
$\left(n+\frac{1}{2}\right)\pi$, corresponding to $S_{00}=-1$. If
there is any inelastic scattering, $|\Gamma_{B0}| < |\Gamma_B|$
and this does not occur. When the circle in $S_{00}$ is small, the
phase shift $\delta_0$ and the scattering length $a$ show only
small peaks or oscillations across a resonance.

The expression (\ref{eqsigela}) for the elastic scattering cross
section saturates at a value $\sigma_{\rm el} \approx 4\pi/k^2$
when $|a|\gg k_0^{-1}$. Such values of $|a|$ occur only when
$\left|\delta_0-\left(n+\frac{1}{2}\right)\pi\right| \ll 1$ and
thus when $\Gamma_B$ is strongly dominated by $\Gamma_{B0}$. Since
$\Gamma_{B0}$ is proportional to $k_0$ and the inelastic
contributions $\Gamma_{Bi}$ are independent of $k_0$, there is a
lower bound on the value of $k_0$ at which this occurs. Denoting
the sum of inelastic contributions to $\Gamma_{B}$ as
$\Gamma_{B}^{\rm inel}$, this is given by
\begin{eqnarray}
|\Gamma_{B}^{\rm inel}| &\ll& |\Gamma_{B0}| = 2k_0 |\gamma_{B0}| \\
k_0 &\gg& \frac{\Gamma_{B}^{\rm inel}}{2\gamma_{B0}}.
\label{eqlimk}
\end{eqnarray}

The radius of the circle in $S_{00}$ is $\Gamma_{B0}/\Gamma_B$.
For small $k_0$, where Eq.\ \ref{eqgamlin} applies, this is
approximately $2k_0 \gamma_{B0}/\Gamma_B^{\rm inel}$. The formula
followed by the complex scattering length is
\begin{equation}
a(B) = a_{\rm bg} + \frac{a_{\rm res}}{2(B-B_{\rm
res})/\Gamma_B^{\rm inel}+{\rm i}}, \label{eqaares}
\end{equation}
where $a_{\rm res}$ is a {\it resonant scattering length} that
characterises the strength of the resonance,
\begin{equation}
a_{\rm res}=\frac{2\gamma_{B0}}{\Gamma_B^{\rm inel}} \, e^{2{\rm
i}(\phi_0+k_0a_{\rm bg})}. \label{eqaresdef}
\end{equation}
Both $a_{\rm res}$ and the background term $a_{\rm bg}$ can in
general be complex and are independent of $k_0$ at low energy. The
phase correction $+2k_0\alpha_{\rm bg}$ in Eq.\ \ref{eqaresdef} is
needed to keep the phase of $a_{\rm res}$ independent of $k_0$.
The explicit expressions for the real and imaginary parts of
$a(B)$ are
\begin{eqnarray}
\alpha(B)=\alpha_{\rm bg} + \frac{\alpha_{\rm res}\left[2(B-B_{\rm
res})/\Gamma_B^{\rm inel}\right] + \beta_{\rm
res}}{\left[2(B-B_{\rm res})/\Gamma_B^{\rm inel}\right]^2+1};\\
\beta(B)=\beta_{\rm bg} + \frac{\alpha_{\rm res} + \beta_{\rm
res}\left[2(B-B_{\rm res})/\Gamma_B^{\rm inel}\right]
}{\left[2(B-B_{\rm res})/\Gamma_B^{\rm inel}\right]^2+1},
\label{eqabres}
\end{eqnarray}
where $a(B)=\alpha(B) - {\rm i}\beta(B)$ and similarly for $a_{\rm
res}$ and $a_{\rm bg}$. The peak profiles for the elastic and
total inelastic cross sections are given by Eqs.\ \ref{eqsigela}
and \ref{eqsiginela}.

In the special case where the background scattering is elastic
($a_{\rm bg}$ is real), unitarity requires that the circle in
$S_{00}$ must loop towards the origin. This requires that $a_{\rm
res}$ is also real. Across the width of the resonance, the real
part $\alpha(B)$ of the scattering length $a(B)$ then oscillates
about $a_{\rm bg}$ by $\pm a_{\rm res}/2$ and the imaginary part
peaks at $\beta(B)=a_{\rm res}$. When the background scattering is
inelastic, however, $a_{\rm res}$ can be complex and the circle in
$S_{00}$ does not point directly towards the origin. The
lineshapes are then unsymmetrical, and $\beta(B)$ (and hence the
inelastic rate) can show a trough as well as a peak. Nevertheless,
the overall magnitude of the oscillations in the scattering length
is still governed by $a_{\rm res}$.

The behaviour derived here is analogous to that observed when
laser light is used to tune scattering lengths
\cite{Fedichev:1996, Bohn:1997}. However, in that case the
amplitude of the oscillation depends on the ratio of excitation
and spontaneous emission rates, which both depend on the same
dipole strength (though the ratio of rates can be tuned with laser
intensity). In the present case $a_{\rm res}$ depends on
independent elastic and inelastic couplings. If $a_{\rm res}$ is
small, the resonant oscillations in cross sections and the
scattering length are small.

The results (\ref{eqaares}) to (\ref{eqabres}) are valid when
$k_0|a_{\rm res}| \ll 1$. Whenever $k_0a_{\rm res} \not\gg 1$,
Eq.\ \ref{eqares} fails at values of $|a|$ small enough to reduce
the height of the peak in the elastic cross section given by Eq.\
\ref{eqsigela}. Conversely, when $k_0|a_{\rm res}| \gg 1$,
$S_{00}$ describes a circle of radius close to 1 in the complex
plane; the behaviour of the scattering length is then well
described by a 2-channel model and the peak in the elastic cross
section is of height $\sim 4\pi/k_0^2$.

The elastic partial width $\Gamma_{B0}$ is proportional to $k_0$
at low energy but becomes constant at high energy. It may be
written \cite{Julienne:2006}
\begin{equation}
\Gamma_{B0}(k_0) = \overline \Gamma_{B0} C_0(k_0)^{-2},
\label{eqgamc}
\end{equation}
where $\overline \Gamma_{B0}$ is independent of $k_0$ and depends
on the short-range coupling between the bound state and the
incoming channel. The factor $C_0(k_0)^{-2}$ is the amplitude
matching function of multichannel quantum defect theory, which is
1 at high energy but near threshold is \cite{Julienne:2006}
\begin{equation}
C_0(k_0)^{-2} = k_0 \overline a \left[1+(1-a_{\rm bg}/\overline
a)^2\right], \label{eqck}
\end{equation}
where $\overline a$ is the mean scattering length
\cite{Gribakin:1993}, $\overline a = 0.478(2\mu
C_6/\hbar^2)^{1/4}$ for a Van der Waals potential $-C_6/r^6$. The
transition between the linear and constant regimes depends on
$C_6$ and the reduced mass \cite{Julienne:1989}, but typically
occurs around $E_{\rm kin}/k_B = 1$~mK.

The height of the peak (or size of the oscillation) in the total
inelastic cross section is proportional to $|a_{\rm res}|$. This
in turn depends principally on the {\it ratio} of
$\overline\Gamma_{B0}$ and $\Gamma_B^{\rm inel}$. Two very
different cases may be distinguished. If the {\em same} coupling
term connects the bound state to the incoming and inelastic
channels, it is likely that $\overline \Gamma_{B0}$ and
$\Gamma_B^{\rm inel}$ will be comparable. Under these
circumstances $a_{\rm res}$ will be of the order of $\overline a$
and there will be relatively small oscillations in the scattering
length. Conversely, if coupling to the inelastic (exoergic)
channels is much weaker than coupling to the elastic channel,
$a_{\rm res}$ will be large and the scattering length will exhibit
a large oscillation resembling a pole.

It is important to realize that $a_{\rm res}$ (and thus the
strength of the resonance) depends on the {\it relative}
magnitudes of the couplings from the {\it resonant state} to the
elastic and inelastic channels. This is {\it not} necessarily the
same as saying that the degree of suppression depends on the
strength of inelastic scattering.

The peaks in {\it individual} inelastic cross sections can be
rather larger than those in $\sigma_{\rm inel}^{\rm tot}$, because
the radius of the circle in $S_{0i}$ is $(2 k_0 |a_{\rm res}|
\Gamma_{Ei} / \Gamma_E^{\rm inel})^{1/2}$, which is considerably
larger than $2k_0 |a_{\rm res}|$ for small $k_0$.

\section{Examples from low-energy atomic and molecular scattering}

For atomic collisions, the couplings to inelastic channels are
sometimes weak enough that a 2-channel model remains accurate even
when inelastic scattering is energetically allowed. For example,
Donley {\em et al.}\ \cite{Donley:2002} and Thompson {\em et al.}\
\cite{Thompson:spont:2005} have produced $^{85}$Rb$_2$ molecules
by magnetic tuning in the vicinity of a Feshbach resonance between
$(f,m_f) = (2,-2)$ states of $^{85}$Rb near 155~G. The $(2,-2)$
state is not the lowest in a magnetic field, and the molecules can
decay by spontaneous spin relaxation to atomic levels with $f=2$
and $m_f>-2$. The resonant state has $M_F=m_{f1}+m_{f2}=-4$, so
this decay requires a change in $M_F$ and involves very weak
magnetic dipole coupling. However, the coupling between the
resonant state and the incoming channel (also $M_F=-4$) is through
much stronger central terms in the potential. K\"{o}hler {\em et
al.}\ \cite{Kohler:2005} have used coupled channel calculations
including spin relaxation to characterize the resonance and
obtained $a_{\rm bg}=-484.1\ a_0$ and $\Delta_B=10.65$~G. Their
lifetime $\tau=32\ \mu$s for the bare resonance state corresponds
to $\Gamma_B^{\rm inel} = \hbar/\tau\Delta\mu = 0.090$~G. With
these parameters, $a_{\rm res} = 1.14 \times 10^5\ a_0$. The
temperature in the experiments of Thompson {\em et al.}\
\cite{Thompson:spont:2005} is 30 nK, corresponding to
$k_0=4.3\times10^{-4}\ a_0^{-1}$. In this system, therefore, $k_0
a_{\rm res}\approx50$ and the resonant behaviour of the scattering
length and the elastic cross section is well approximated by a
2-channel model.

\begin{figure}[tb]
\begin{center}
\includegraphics[width=0.7\linewidth]{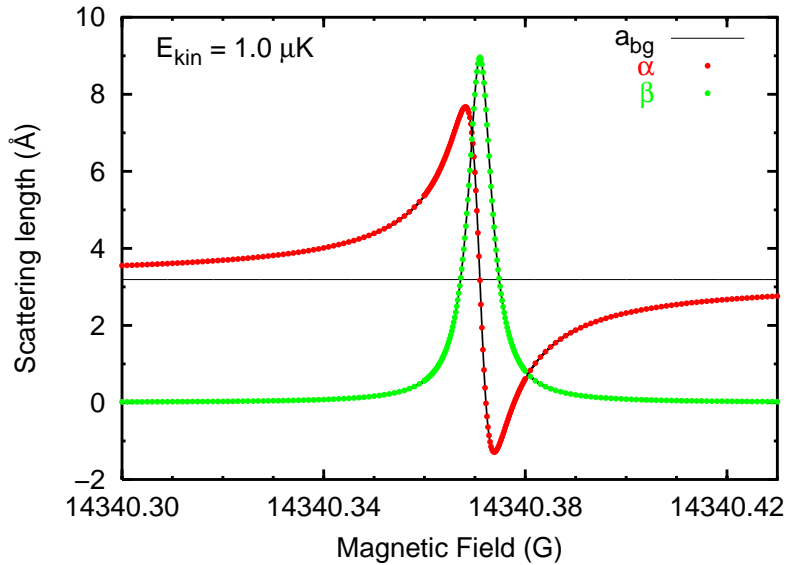}
\end{center}
\caption{Real (red) and imaginary (green) parts of the scattering
length for $^3$He + NH collisions in the vicinity of an inelastic
Feshbach resonance at a kinetic energy of $10^{-6}$~K. The lines
show the results of Eq.\ \ref{eqaares}. This is the same resonance
as shown in the right-hand panel of Figure \ref{figsm}.}
\label{aHeNH}
\end{figure}

The situation is very different for rotationally inelastic
molecular scattering, where the potential anisotropy couples the
resonant bound state to both the incoming and inelastic channels.
Under these circumstances $a_{\rm res}$ will generally be small.
In separate work, we have described numerical tests of the
equations derived here for He + NH($^3\Sigma^-$) scattering in a
magnetic field \cite{Gonzalez-Martinez:2007}. This is a very
weakly coupled system, and for the rotational ground state ($n=0$)
of NH the channels with different spin projections $m_s$ are
coupled only indirectly via excited rotational levels. The
background scattering is essentially elastic, so $a_{\rm bg}$ and
$a_{\rm res}$ are real. Fig.\ \ref{aHeNH} shows the real and
imaginary parts of the scattering length for magnetic tuning
across an inelastic scattering Feshbach resonance in this system.
Even He + NH, where the inelastic couplings are much weaker than
in most other molecular systems, $a_{\rm res}\approx 9$~\AA\ and
$k_0 a_{\rm res}\ll 1$. The oscillations in scattering lengths and
elastic cross sections are strongly suppressed at low energies.

There are also atomic systems where the coupling to inelastic
channels is strong enough to suppress the oscillations in
scattering lengths. Such effects have been observed, for example,
in calculations on collisions of Sr ($^2P_2$) atoms
\cite{Kokoouline:2003}, where the bound state is coupled to both
the incoming and inelastic channels by anisotropic potential
terms.

Eqs.\ (\ref{eqaares}) to (\ref{eqabres}) can be adapted to apply
to any parameter $\lambda$ that tunes scattering resonances across
a threshold. The ratio $\Gamma_{\lambda0} / \Gamma_\lambda^{\rm
inel}$ is the same for any such parameter (and is equal to
$\Gamma_{E0}/\Gamma_E^{\rm inel}$). The resonant scattering length
therefore has the same value for any parameter $\lambda$. $a_{\rm
res}$ is a universal measure of the strength of a low-energy
resonance, independent of the parameter used to tune it through a
threshold.

This explains previously puzzling results obtained in low-energy
reactive scattering. Qu\'{e}m\'{e}ner {\em et al.}\
\cite{Quemener:2004} and Cvita\v{s} {\em et al.}\
\cite{Cvitas:li3:2007} have investigated the sensitivity of
scattering cross sections in Na + Na$_2$ and Li + Li$_2$ to
variations in the potential energy surface. Scaling the potential
tunes reactive scattering resonances across threshold, and this
produces oscillations in the elastic and inelastic cross sections.
In these systems the couplings to individual vibrationally
inelastic channels are somewhat reduced by the large kinetic
energy release, so that for low initial $v$ (with relatively few
inelastic channels) some significant resonant peaks remain. For
initial $v=1$, the cross sections oscillate by about a factor of
10 as resonances cross threshold. Even this corresponds to a
relatively small oscillation in the complex scattering length
(small $a_{\rm res}$). However, the amplitudes of the oscillations
decrease substantially with increasing vibrational excitation of
the colliding molecules and are almost smooth for $v=3$ for both
Na + Na$_2$ \cite{Quemener:2004} and Li + Li$_2$
\cite{Cvitas:li3:2007}.

Quite different behaviour has been observed in F + H$_2$ reactions
\cite{Bodo:FH2:2004}, but is also explained by the present theory.
Bodo {\em et al.}\ \cite{Bodo:FH2:2004} investigated the effect of
scaling the reduced mass and observed pole-like behaviour in the
scattering length and large reactive cross sections as a resonance
was tuned across threshold. In this case the resonant state is
localised in the entrance channel of the reaction, while the only
exoergic channels are reactive ones that are separated from the
entrance channel by a high barrier. $\Gamma_E^{\rm inel}$ is thus
reduced relative to $\overline \Gamma_{E0}$. Because of this,
$a_{\rm res}$ is large ($>$100~\AA) and no strong suppression of
the resonant peaks occurs.

The considerations of the present paper lead to a remarkable
conclusion. It has been commonly believed that collision cross
sections in the ultracold regime are extremely sensitive to
details of the potential energy surface, and that for molecules
these dependences would be even more limiting than for atoms. The
present paper has shown that this is true only when the resonant
state is coupled much more weakly to inelastic (exoergic) channels
than to the incoming channel. There are some systems where
inelastic processes are weak enough for scattering lengths to
reach near-infinite values at zero-energy resonances. However, in
other cases inelastic processes will suppress this behaviour. In
general terms, the resonant peaks are suppressed by inelastic
scattering unless there is a specific mechanism that reduces the
coupling to inelastic channels.

\section*{Acknowledgments}

The author is grateful to Paul Julienne, Maykel Leonardo
Gonz{\'a}lez-Mart{\'\i}nez and Marko Cvita\v s for comments on the
manuscript.

\bigskip
\bibliographystyle{aip}
\bibliography{all}

\end{document}